\documentclass[12pt]{amsart}
\usepackage{amsmath}
\usepackage{graphicx}
\usepackage{amsfonts}
\usepackage{amssymb}
\newtheorem{theorem}{Theorem}

\newtheorem{proposition}[theorem]{Proposition}

\begin{document}
\author{Nolan R. Wallach}
\title{An unentangled Gleason's theorem}
\maketitle

\begin{abstract}
The purpose of this note is to give a generalization of Gleason's theorem
inspired by recent work in quantum information theory. For multipartite
quantum systems, each of dimension three or greater, the only nonnegative
frame functions over the set of unentangled states are those given by the
standard Born probability rule. However, if one system is of dimension 2
this
is not necessarily the case.

\end{abstract}

\section{Introduction.}

Let $H$ be a Hilbert space with unit sphere $S(H)$. Following Gleason
([Gleason]) we will call a function $f:S(H)\rightarrow\mathbb{C}$ a frame
function of weight $w$ if for every orthonormal basis $\{v_{i}\}$ of $S(H)$
\begin{equation}
\sum_{i}f(v_{i})=w\text{.} \label{wght}%
\end{equation}
In [Gleason] the following theorem was proved
\bigskip

\begin{theorem}
If $\dim H\geq3$ and $f$ is a frame function that takes non-negative real
values then there exists a self adjoint trace class operator $T:H\rightarrow
H$ such that
\[
f(v)=\langle v|T|v\rangle,v\in S(H)\text{.}%
\]
\end{theorem}

\bigskip

This theorem is of importance to quantum mechanics because it allows a
significant weakening of the axioms, showing that the Born probability rule
[Born] provides the unique class of probability assignments for measurement
outcomes so long as those probabilities are specified by frame functions
[Pitowsky]. The theorem also rules out a large class of hidden-variable
explanations for quantum statistics, the so-called noncontextual hidden
variables, in dimension $3$ or greater. The interested reader should consult
[Bell] for a discussion of this point. If the Hilbert space is of dimension
$2$, then the statement in the theorem is easily seen to be false.

The purpose of this note is to give a generalization of Gleason's theorem
inspired by recent work in quantum information theory. In that context the
issue of \emph{local} measurements and operations on multipartite quantum
systems (as opposed to the full set of operations) is of the utmost
importance
[BDFMRSSW]. For instance, it has been pointed out that probabilities for the
outcomes of local measurements are enough to uniquely specify the quantum
state from which they arise if the field of the Hilbert space is complex,
though this fails for real and quaternionic Hilbert spaces [Araki,Wootters].

Chris Fuchs has asked to what extent local and semi-local measurements not
only uniquely specify the quantum state, but also a Born-like rule as in
Gleason's result [Fuchs]. In this regard, the following formalization
appears
natural. We confine our attention to finite dimensional Hilbert spaces for
the
sake of simplicity. Let $H_{1},...,H_{n}$ be Hilbert spaces. Set
$H=H_{1}\otimes H_{2}\otimes\cdot\cdot\cdot\otimes H_{n}$. Let $\Sigma
=\Sigma(H_{1},...,H_{n})$ denote the subset of $S(H)$ consisting of those
elements of the form $a_{1}\otimes\cdot\cdot\cdot\otimes a_{n}$ with
$a_{i}\in
S(H_{i})$ for $i=1,...,n$. In the jargon of quantum information theory such
states are called \emph{unentangled} or \emph{product states}. The ones that
are not of this form are said to be \emph{entangled}. An orthonormal basis
$\{v_{i}\}$ of $H$ is said to be unentangled if $v_{i}\in\Sigma$ for all
$i$.
We say that $f:\Sigma\rightarrow\mathbb{C}$ is an unentangled frame function
of weight $w$ if whenever $\{v_{i}\}$ is an unentangled orthonormal basis of
$H$ then $f$ satisfies (1) above. We establish the following result.

\bigskip

\begin{theorem}
If $\dim H_{i}\geq3$ for all $i\,$\ and if $f:\Sigma\rightarrow\mathbb{R}$
is
a non-negative unentangled frame function then there exists $T:H\rightarrow
H$
a self adjoint trace class operator such that $f(v)=\langle v|T|v\rangle$
for
all $v\in\Sigma$.
\end{theorem}

\bigskip

This theorem is an almost direct consequence of Gleason's original theorem.
We
will give a proof of it in the next section. The second result in this paper
shows that the dimensional condition is necessary.

It should be noted however, that despite the absence of entangled or
``nonlocal'' states in $\Sigma$, in [BDFMRSSW] it is asserted that not all
unentangled bases correspond to quantum measurements that can be carried out
by local means alone (even with iterative procedures based on weak local
measurements and unlimited amounts of classical communication between the
measurers at each site). The simplest kind of purely local measurement is
given by an alternative type of basis adapted to the tensor product
structure.
This is a \emph{product basis} and is defined as to be a basis of the form
$\{u_{i_{1}1}\otimes u_{i_{2}2}\otimes\cdot\cdot\cdot\otimes u_{i_{n}n}\}$
where $u_{1j},...,u_{n_{j}j}$ is an orthonormal basis of $H_{j}$. We could
define a product frame function in the same way as we did for an unentangled
frame function except that we only assume that there exists a weight $w$
such
that $\sum_{i_{1},i_{2},...,i_{n}}f(u_{i_{1}1}\otimes u_{i_{2}2}\otimes
\cdot\cdot\cdot\otimes u_{i_{n}n})=w$ for every product basis. One can ask
whether this is all that us necessary for the conclusion of the theorem
above.
The answer is no and a method of ``finding'' a large class of examples will
be
given at the end of the next section (see the proposition at the end of the
section). This result amasses some evidence that the structure of local
measurements alone is not enough to establish the Born rule for multipartite
systems, but a full answer would require consideration of the largest class
of
local measurements in [BDFMRSSW].

These issues also spawn another theorem.

\bigskip

\begin{theorem}
Let $\dim H_{1}=2$ and let $f:S(H_{1})\rightarrow\mathbb{C}$ be a frame
function of weight $w_{1}$ and $g:\Sigma(H_{2},...,H_{n})\rightarrow
\mathbb{C}$ be an unentangled frame function of weight $w_{2}$. We set
$h(v_{1}\otimes u)=f(v_{1})g(u)$ for $u\in\Sigma(H_{2},...,H_{n})$. Then $h$
is an unentangled frame function of weight $w_{1}w_{2}$.
\end{theorem}

\bigskip

This result is a bit harder and the proof involves a method (see Theorem 5)
that describes a combinatorial scheme for finding all unentangled
orthonormal
bases where all of the spaces, $H_{i}$, have dimension $2$. This analysis in
turn leads to a natural question. Given and unentangled orthonormal set can
it
be extended to an unentangled orthonormal basis? Or even stronger: Can it be
a
proper subset of an unentangled orthonormal set? This question was studied
in
[BDMSST]. We conclude the paper by giving a proof based on simple algebraic
geometry of the following theorem which is related to the bound that occurs
in [BDMSST].

\bigskip

\begin{theorem}
Let $V$ be a subspace of $H_{1}\otimes\cdot\cdot\cdot\otimes H_{n}$ such
that
if $v\in V$ and $v\neq0$ then $v$ is entangled. Then $\dim V\leq\dim
(H_{1})\cdot\cdot\cdot\dim(H_{n})-\sum(\dim H_{i}-1)-1$. Furthermore, the
upper bound is attained.
\end{theorem}

\bigskip

\section{The unentangled Gleason theorem.}

In this section we will give a proof of Theorem 2. If $n=1$ the statement is
just Gleason's theorem. We consider the situation of $H=H_{0}\otimes V$ with
$V=H_{1}\otimes H_{2}\otimes\cdot\cdot\cdot\otimes H_{n}$ and $\dim H_{i}%
\geq3$ for all $i$. We prove Theorem 1 by induction (i.e. assume the result
for $n)$. We note that if $\{v_{i}\}$ is an orthonormal basis of $H_{0}$ and
if for each $i$, $\{u_{ij}\}$ is an unentangled orthonormal basis of $V$
then
the set $\{v_{i}\otimes u_{ij}\}$ is an unentangled orthonormal basis of
$H$.
Thus if $w$ is the weight of $f$ then we have
\[
\sum_{j}f(v_{1}\otimes u_{1j})=w-\sum_{i\geq2,j}f(v_{i}\otimes u_{ij}).
\]
Thus for each $v\in S(H_{0})$ the function $f_{v}(u)=f(v\otimes u)$ is an
unentangled frame function. The inductive hypothesis implies that for each
$v\in S(H_{0})$ there exists a self adjoint (due to the reality of $f$)
linear
operator $T(v)$ such that $f(v\otimes u)=\langle u|T(v)|u\rangle$ for
$u\in\Sigma(H_{1},...,H_{n})$. Similarly, if $\{u_{i}\}$ is an unentangled
orthonormal basis of $V$ and for each $i$, $\{v_{ij}\}$ is an orthonormal
basis of $H_{0}$ then $\{v_{ij}\otimes u_{i}\}$ is an unentangled
orthonormal
basis of $H$. We therefore conclude as above that if $u\in\Sigma
(H_{1},...,H_{n})$ then there exists $S(u)$ a self adjoint linear operator
on
$H_{0}$ so that $f(v\otimes u)=\langle v|S(u)|v\rangle$ for all $v\in
H_{0}$.

Let $\{u_{i}\}$ be an unentangled orthonormal basis of $V$ and let
$\{v_{j}\}
$ be an orthonormal basis of $H_{0}$. Set
\[
a_{ij}(v)=\langle u_{i}|T(v)|u_{j}\rangle
\]
and
\[
b_{ij}(u)=\langle v_{i}|S(u)|v_{j}\rangle,u\in\Sigma(H_{1},...,H_{n}).
\]
We now observe that if $v=\sum_{i}x_{i}v_{i}$ and if $u=\sum_{j}y_{j}u_{j}$
then we have
\[
\sum_{p,q}a_{p,q}(v)\bar{y}_{p}y_{q}=\sum_{r,s}b_{r,s}(u)\bar{x}_{r}x_{s }.
\]
If we substitute $v=v_{r}$ then we have
\[
b_{rr}(u)=\sum_{p,q}a_{p,q}(v_{r})\bar{y}_{p}y_{q}\text{.}%
\]
Now assuming that $r\neq s$ and taking $v=\frac{1}{\sqrt{2}}(v_{r}+v_{s})$
we
have
\begin{align*}
\operatorname{Re}b_{rs}(u)  &  =\sum_{p,q}a_{p,q}(\frac{1}{\sqrt{2}}%
(v_{r}+v_{s}))\bar{y}_{p}y_{q}-\\
&  \frac{1}{2}\left(  \sum_{p,q}a_{p,q}(v_{r})\bar{y}_{p}y_{q}+\sum
_{p,q}a_{p,q}(v_{s})\bar{y}_{p}y_{q}\right)  .
\end{align*}
Also if we take $v=\frac{1}{\sqrt{2}}(v_{r}+iv_{s})$ then we have
\begin{align*}
-\operatorname{Im}b_{rs}(u)  &  =\sum_{p,q}a_{p,q}(\frac{1}{\sqrt{2}}%
(v_{r}+iv_{s}))y_{p}\overline{y}_{q}-\\
&  \frac{1}{2}\left(  \sum_{p,q}a_{p,q}(v_{r})y_{p}\overline{y}_{q}+\sum
_{p,q}a_{p,q}(v_{s})y_{p}\overline{y}_{q}\right)  .
\end{align*}

Thus if we set
\[
c_{rrpq}=a_{pq}(v_{r})
\]
and if $r\neq s$ then
\begin{align*}
c_{rspq}  &  =a_{p,q}\left(  \frac{1}{\sqrt{2}}(v_{r}+v_{s})-\frac{1}%
{2}(a_{p,q}(v_{r})+a_{p,q}(v_{s}))\right)  +\\
&  a_{p,q}\left(  \frac{1}{\sqrt{2}}(v_{r}+iv_{s})-\frac{1}{2}(a_{p,q}%
(v_{r})+a_{p,q}(v_{s}))\right)
\end{align*}
Then
\[
f(v\otimes u)=\sum_{rspq}c_{rspq}\bar{x}_{r}\bar{y}_{p}x_{s}y_{q}.
\]
This is the content of the theorem.

\bigskip

We will now give a counterexample to the analogous assertion for product
bases.

\begin{proposition}
Let $H_{1}$ and $H_{2}$ be finite dimensional Hilbert spaces of dimension
greater than $1$. Then there exists $f:\Sigma(H_{1},H_{2})\rightarrow
\lbrack0,\infty)$ such that $\sum_{i,j}f(u_{i}\otimes v_{j})=w$, with
$w\in\mathbb{R}$ fixed, for all choices $\{u_{i}\}$ and $\{v_{j}\}$ of
orthonormal bases of $H_{1}$ and $H_{2}$ respectively but there is no linear
endomorphism, $T$, on $H_{1}\otimes H_{2}$ such that $f(u\otimes
v)=\left\langle u\otimes v|T|u\otimes v\right\rangle $ for $u\in S(H_{1})$
and
$v\in S(H_{2})$.
\end{proposition}

\bigskip

\noindent Proof. Let for $w>0$, $\mathcal{P}_{w}$ denote the set of all
Hermitian positive semi-definite endomorphisms, $A$, of $H_{2}$ such that
$tr(A)=w$. \ Fix $w_{o}=\frac{w}{\dim H_{1}}$. Let $\varphi:S(H_{1}%
)\rightarrow\mathcal{P}_{w_{o}}$ be a mapping (completely arbitrary). Set
$f(u\otimes v)=\left\langle v|\varphi(u)|v\right\rangle $, for $u\in
S(H_{1})$
and $v\in S(H_{2})$ . If $\{u_{i}\}$ is an orthonormal basis of $H_{1}$ and
if
$\{v_{j}\}$ is an orthonormal basis of $H_{2}$ then
\[
\sum_{i,j}f(u_{i}\otimes v_{j})=\sum_{i}\left(  \sum_{j}\left\langle
v_{j}|\varphi(u_{i})|v_{j}\right\rangle \right)  =\sum_{i}tr(\varphi
(u_{i}))=\dim(H_{1})w_{o}.
\]

\bigskip

\noindent\textbf{Note}: In this argument only one factor need be finite
dimensional. Also note that $f$ can be chosen to be continuous.

\section{Unentangled Bases}

In this section we will develop the material on ``unentangled bases'' that
we
will need to prove Theorem 3 (in fact as we shall see a generalization). Let
$V$ be a $2$-dimensional Hilbert space and let $H$ be an $n$-dimensional
Hilbert space. Fix $\Sigma\subset S(H)$ such that $\lambda\Sigma=\Sigma$ for
all $\lambda\in\mathbb{C}$ with $|\lambda|=1$. We will use the notation
$S(V)\otimes\Sigma=\{v\otimes w|v\in S(V),w\in\Sigma\}$.

If $a\in S(V)$ then up to scalar multiple there is exactly one element of
$S(V)$ that is perpendicular to $a$. We will denote a choice of such an
element by $\widehat{a}$. The main result of this section is

\bigskip

\begin{theorem}
If $\{u_{j}\}_{j=1}^{2n}$ is an orthonormal basis of $V\otimes H$ with
$u_{j}\in S(V)\otimes\Sigma$ for $j=1,...,2n$ then there exists a partition
\[
n_{1}\geq n_{2}\geq...\geq n_{r}>0
\]
of $n$, an orthogonal decomposition
\[
H=U_{1}\oplus...\oplus U_{r},
\]
elements $a_{1},...,a_{r}\in S(H)$, and for each $i=1,...,r$ orthonormal
bases
$\{b_{i1},...,b_{in_{i}}\}$ and $\{c_{i1},...,c_{in_{i}}\}$ of $U_{i}$ such
that
\[
\{u_{i}|\,i=1,...,2n\}=\bigcup_{i=1}^{r}\left(  \{a_{i}\otimes b_{ij}%
|\,j=1,...,n_{i}\}\cup\{\widehat{a}_{i}\otimes c_{ij}|\,j=1,...,n_{i}%
\}\right)  .
\]
\end{theorem}

\bigskip

Before we prove the theorem we will make several preliminary observations.
Let
$\{u_{i}\}$ be as in the statement of the theorem. Then each $u_{i}%
=a_{i}\otimes h_{i}$ with $a_{i}\in S(V)$ and $h_{i}\in\Sigma$.

\bigskip

\noindent1. For each $i$ there exists $j$ such that $a_{j}$ is a multiple of
$\widehat{a}_{i}$.\smallskip

If not then we would have $\langle a_{i}|a_{j}\rangle\neq0$ for all $j$.
Since
$\langle a_{i}\otimes h_{i}|a_{j}\otimes h_{j}\rangle=\langle a_{i}
|a_{j}\rangle\langle h_{i}|h_{j}\rangle$, $\langle h_{i}|h_{j}\rangle=0$ for
all $j\neq i$. This implies that $\{u_{j}\}_{j\neq i}\subset V\otimes
\{h_{i}^{\perp}\}$. This space has dimension equal to $2(n-1)$. So it could
not contain $2n-1$ orthonormal elements. This contradiction implies that
assertion 1. is true.\smallskip

\noindent2. Assume that $i\neq j$. If $\langle a_{i}|a_{j}\rangle\neq0$ then
$\langle h_{i}|h_{j}\rangle=0$. If $\langle h_{i}|h_{j}\rangle\neq0$ then
$\langle a_{i}|a_{j}\rangle=0$.\smallskip\

This is clear (see the proof of 1.)

\bigskip

We will now prove the theorem by induction on $n$. If $n=1$ the result is
trivial. We assume the result for all $H$ with $\dim H<n$ and all possible
choices for $\Sigma$. We now prove it for $n$.

For each $i$ let $m_{i}$ denote the number of $j$ such that $a_{j}$ is a
multiple of $a_{i}$. Let $m=\max\{m_{i}|i=1,...,2n\}$. If we relabel we may
assume that the first $m$ of the $a_{i}$ are equal to $a_{1}$ (we may have
to
multiply $h_{i}$ by a scalar of norm $1$). By 1. above we may assume that
the
next $k$ of the $a_{i}$ are equal to $\widehat{a}_{1}$ with $1\leq k\leq m$
and if $i>m+k$ then $a_{i}$ is not a multiple of either $a_{1}$ or
$\widehat{a}_{1}$. This implies by 2. above that $\langle
h_{i}|h_{j}\rangle=0
$ for $j>m+k$ and $i=1,...,m$. Also $\{h_{1},...,h_{m}\}$ is an orthonormal
set. Thus $u_{i}\in V\otimes(\{h_{1},...,h_{m}\}^{\perp})$ for $i>m+k$. This
implies that $V\otimes(\{h_{1},...,h_{m}\}^{\perp})$ contains $2n-(m+k)$
orthonormal elements. Since $\dim V\otimes(\{h_{1},...,h_{m}\}^{\perp
})=2(n-m)$ this implies that $k=m$. We now rewrite the first $2m$ elements
of
the basis as
\[
a_{1}\otimes b_{1},...,a_{1}\otimes b_{m},\widehat{a}_{1}\otimes
c_{1},...,\widehat{a}_{1}\otimes c_{m}.
\]
If we apply observation 2. again we see that the elements $h_{i}$ for $i>2m$
must be orthogonal to $\{b_{1},...,b_{m}\}$ and to $\{c_{1},...,c_{m}\}$. A
dimension count says that they must span the orthogonal complements of both
$\{b_{1},...,b_{m}\}$ and $\{c_{1},...,c_{m}\}$. But then
$\{b_{1},...,b_{m}\}
$ and $\{c_{1},...,c_{m}\}$ must span the same space, $U\subset H$. We have
therefore shown that $\{u_{i}\}_{i>2m}$ is an orthonormal basis of $V\otimes
U^{\perp}$. We may thus apply the inductive hypothesis to $U^{\perp}$ and
$\Sigma\cap U^{\perp}$. This completes the inductive step and hence the
proof.

\bigskip

If $W$ is a Hilbert space and if $\Xi$ is a subset of $S(W)$ that is
invariant
under multiplication by scalars of absolute value $1$ then a function
$f:\Xi\rightarrow\mathbb{C}$ is said to be a $\Xi$-frame function of weight
$w=w_{f}$ if whenever $\{u_{i}\}$ is an orthonormal basis of $W$ with
$u_{i}\in\Xi$ (i.e. $\{u_{i}\}$ is a $\Xi$-frame) we have $\sum_{i}f(u_{i}%
)=w$. We note

\bigskip

\noindent3. Let $f$ be a $\Xi$-frame function. If $\{u_{i}\}$ is a
$\Xi$-frame
for $W$ and if $F$ is a subset of $\{u_{i}\}$ then $f_{|F^{\perp}\cap\Xi}$
is
a $F^{\perp}\cap\Xi$-frame function of weight $w_{f}-\sum_{u_{i}\in F}
f(u_{i})$.\smallskip

This is pretty obvious. Let $\{v_{j}\}$ be a $\Xi\cap F^{\perp}$-frame for
$F^{\perp}$. Then $\{\nu_{j}\}\cup F$ is a $\Xi$-frame for $W$.

\bigskip

\begin{proposition}
Let $V$ be a two dimensional Hilbert space and let $H$ be an $n$-dimensional
Hilbert space. Let $\Sigma\subset S(H)$ be as in the rest of this section
and
let $g:S(V)\rightarrow\mathbb{C}$ and $h:\Sigma\rightarrow\mathbb{C}$ be
respectively a frame function and a $\Sigma$-frame function. Then if
$f(v\otimes w)=g(v)h(w)$ for $v\in S(H)$ and $w\in\Sigma$ then $f$ is an
$S(V)\otimes\Sigma$-frame function of weight $w_{g}w_{h}$.
\end{proposition}

\bigskip

\noindent Proof. Let $\{u_{i}\}$ be an $S(V)\otimes\Sigma$-frame. Then
Theorem
5 implies that we may assume that there is partition $n_{1}\geq n_{2}
\geq...\geq n_{r}>0$ of $n$ and elements $a_{i},b_{ij}\,$and $c_{ij}$ as in
the statement so that
\[
\{u_{i}\}=\bigcup_{i=1}^{r}\left(  \{a_{i}\otimes b_{ij}|j=1,...,n_{i}%
\}\cup\{\widehat{a}_{i}\otimes c_{ij}|j=1,...,n_{i}\}\right)  .
\]
Thus
\[
\sum_{i}f(u_{i})=\sum_{i}g(a_{i})\sum_{j=1}^{n_{i}}h(b_{ij})+\sum
_{i}g(\widehat{a}_{i})\sum_{j=1}^{n_{i}}h(c_{ij}).
\]
Observation 3. above implies that for each $i$ we have $\sum_{j=1}^{n_{i}%
}h(b_{ij})=\sum_{j=1}^{n_{i}}h(c_{ij})$. Now $g(a_{i})+g(\widehat{a}
_{i})=w_{g}$. Hence since $\{b_{ij}\}$ is a $\Sigma$-frame the result
follows.

\bigskip

Theorem 3 is an immediate consequence of the above proposition.

\bigskip

\section{Entangled subspaces.}

Let $H_{1},...,H_{n}$ be finite dimensional Hilbert spaces and set
$H=H_{1}\otimes H_{2}\otimes\cdot\cdot\cdot\otimes H_{n}$. If $V\subset H$
is
a subspace than we will say then $V$ is \emph{entangled} if whenever $v\in
V$
and $v\neq0$ then $v$ is entangled (i.e. $v$ cannot be written in the form
$v=h_{1}\otimes h_{2}\otimes\cdot\cdot\cdot\otimes h_{n}$ for any choice of
$h_{i}\in H_{i}$). The purpose of this section is to give a proof of Theorem
4
using basic algebraic geometry. That is, we will prove that
\[
\dim V\leq\dim(H_{1})\cdot\cdot\cdot\dim(H_{n})-\sum(\dim H_{i}-1)-1
\]
and that this estimate is best possible. The reader should consult
[Hartshorne] for the algebraic geometry used in the proof of this result.

Let $L=\{\lambda\in H^{\ast}|\,\lambda(V)=0\}$ ($H^{\ast}$ the complex dual
space of $H$). Let $X=\{h_{1}\otimes\cdot\cdot\cdot\otimes
h_{n}\,|\,h_{i}\in
H_{i}\}$. We consider the map $\Phi:H_{1}\times...\times H_{n}\rightarrow X$
given by $\Phi(h_{1},...,h_{n})=h_{1}\otimes\cdot\cdot\cdot\otimes h_{n}$.
Then $\Phi$ is a surjective polynomial mapping. If we denote by $\overline
{\Phi}$ the corresponding mapping of projective spaces we have $\overline
{\Phi}:P(H_{1})\times...\times P(H_{n})\rightarrow P(H)$. General theory
implies that the image of $\overline{\Phi}$ is Zariski closed in $P(H)$.
Since
$X$ is clearly the cone on that image we see that $X$ is Zariski closed and
irreducible. Also the map $\overline{\Phi}$ is injective so the dimension
over
$\mathbb{C}$ of its image is $\sum(\dim H_{i}-1)$. Thus the dimension over
$\mathbb{C}$ of $X$ is $d=\sum(\dim H_{i}-1)+1$.

Since $V$ is entangled $X\cap V=\{0\}$. This implies that $\{x\in
X\,||\lambda(x)=0,\lambda\in L\}=\{0\}$. Thus $\dim L\geq\dim X=d$. Hence
$\dim V=\dim H-\dim L\leq\dim H-d$. This is the asserted upper bound. The
fact
that this upper bound is best possible follows from the Noether
normalization
theorem which implies that there exist $\lambda_{1},...,\lambda_{d}\in
H^{\ast}$ such that $\{x\in X\,||\lambda_{i}(x)=0$ for all $i\}=\{0\}$ (i.e.
a
linear system of parameters).

\bigskip

\begin{center}
{\LARGE References}
\end{center}

$\bigskip$

\noindent\lbrack Gleason] A. M. Gleason, Measures on the closed subspaces of
a
Hilbert space, J. Math. and Mech. 6 (1957), 885--893.

\bigskip

\noindent\lbrack Born] M. Born, Zur Quantenmechanik der Stossvorg\"ange,
Zeits.\ Phys.\ 37 (1926), 863--867. Reprinted and translated in
\emph{Quantum
Theory and Measurement}, edited by J. A. Wheeler and W. H. Zurek (Princeton
U.
Press, Princeton, NJ, 1983), pp.~52--55.

\bigskip

\noindent\lbrack Pitowsky] I.~Pitowsky, Infinite and finite Gleason's
theorems
and the logic of indeterminacy, J. Math.\ Phys.\ 39 (1998), 218--228.

\bigskip

\noindent\lbrack Bell] J. S. Bell, On the problem of hidden variables in
quantum mechanics, Rev.\ Mod.\ Phys.\ 38 (1966), 447--452.

\bigskip

\noindent\lbrack BDFMRSSW] C. H. Bennett, D. P. DiVincenzo, C. A. Fuchs, T.
Mor, E. Rains, P. W. Shor, J. A. Smolin, and W. K. Wootters, Quantum
nonlocality without entanglement, Phys.\ Rev.\ A 59 (1999), 1070--1091.

\bigskip

\noindent\lbrack Araki] H. Araki, On a characterization of the state space
of
quantum mechanics, Comm.\ Math.\ Phys 75 (1980), 1--24.

\bigskip

\noindent\lbrack Wootters] W. K. Wootters, Local accessibility of quantum
states, in \emph{Complexity, Entropy and the Physics of Information}, edited
by W. H. Zurek (Addison-Wesley, Redwood City, CA, 1990), pp.~39--46.

\bigskip

\noindent\lbrack Fuchs] C. A. Fuchs, private communication.

\bigskip

\noindent\lbrack BDMPSST] C. H. Bennett, D. P. DiVincenzo, T.Mor, P. W.
Shor,
J. A. Smolin, B. M. Terhal, Unextendible product bases and bound
entanglement,
Phys.\ Rev.\ Lett.\ 82 (1999) 5385--5388.

\bigskip

\noindent\lbrack Hartshorne] R. Hartshorne, \emph{Algebraic Geometry},
Graduate Texts in Mathematics, 52, Springer-Verlag, New York, 1977.

\bigskip

\textsc{Nolan R. Wallach}

University of California, San Diego

\emph{E-mail address:} nwallach@ucsd.edu
\end{document}